\documentclass{aa}
\usepackage{graphics}
\usepackage{times}

\newcommand{\grs}   {GRS 1758$-$258}
\newcommand{\unoe}  {1E 1740.7$-$2942}

\newcommand{\grp}   {${\rlap.}^{\circ}$}
\newcommand{\pri}   {${\rlap.}^{\prime \prime}$}
\newcommand{\rl}    {${\rlap.}^{s}$}

\newcommand{\ltsima} {$\; \buildrel < \over \sim \;$}
\newcommand{\simlt}  {\lower.5ex\hbox{\ltsima}}            
\newcommand{\gtsima} {$\; \buildrel > \over \sim \;$}
\newcommand{\simgt}  {\lower.5ex\hbox{\gtsima}}            

\begin{document}

\thesaurus{06(08.09.2 \object{GRS 1758$-$258}; 13.25.5; 13.18.5)}

\title
{Search for the optical and infrared counterpart of GRS 1758$-$258}

\author{J. Mart\'{\i}\inst{1,2}
\and S. Mereghetti\inst{3}
\and S. Chaty\inst{1}
\and I.F. Mirabel\inst{1,4}
\and P. Goldoni\inst{1}
\and L.F. Rodr\'{\i}guez\inst{5}
}

\institute{
DAPNIA/Service d'Astrophysique, CEA/Saclay,
F-91191 Gif-Sur-Yvette, France
\and
Departamento de F\'{\i}sica Aplicada, 
Escuela Polit\'ecnica Superior, Universidad de Ja\'en,
Calle Virgen de la Cabeza, 2, E-23071 Ja\'en, Spain
\and
Istituto di Fisica Cosmica G.Occhialini, 
via Bassini 15, I-20133 Milano, Italy
\and
Instituto de Astronom\'{\i}a y F\'{\i}sica del Espacio, C.C. 67, Suc. 28, 1428 Buenos Aires,
Argentina
\and
Instituto de Astronom\'{\i}a, UNAM, Apdo. Postal 70-264,
04510 M\'exico D.F., Mexico}

\offprints{J. Mart\'{\i}, jmarti@discovery.saclay.cea.fr}

\date{Received / Accepted}

\authorrunning{Mart\'{\i} et al.}
\titlerunning{Search for the optical and infrared counterpart of GRS 1758$-$258}
\maketitle

\begin{abstract}
We report the results of a deep search for the optical and near infrared
counterpart of the microquasar source \grs. 
At least two possible candidate counterparts of the binary star companion
have been recognized on the basis of astrometric coincidence
to within $1^{\prime\prime}$. Our photometric study shows that the 
brightest of them would be consistent with a K-type giant star, while the
weakest one would be a main sequence F companion. Follow up spectroscopic observations 
in the near infrared $H$ and $K$-bands have failed so far to provide evidence for emission
lines that may support an unambiguous identification. 
However, the proximity of these two
sources to the sub-arcsec VLA radio position of \grs\ makes them deserving further 
attention in the future. 


\end{abstract}

\section{Introduction}

\grs\ is one of the two persistent hard X-ray ($\geq30$ keV) emitters in the Galactic Center region
together with \unoe\ (Sunyaev et al. 1991;   Goldwurm et al. 1994). The two sources are  
known to exhibit radio counterparts with double-sided jets emanating from a 
central compact source (Mirabel et al. 1992a; Rodr\'{\i}guez, Mirabel \& Mart\'{\i} 1992).
This morphological analogy with extragalactic AGNs and quasars is part of the motivation
for considering both \grs\ and \unoe\ to be members of the microquasar class of 
black hole galactic X-ray binaries 
(see e.g. Mirabel \& Rodr\'{\i}guez 1998 for a recent review). However, 
the   galactic origin of \grs\ and \unoe\ has not been
verified yet by means of classical photometric or spectroscopic studies of their
optical/infrared counterpart, provided of course that we could actually detect it.  

For a successful counterpart search in the crowded fields of the Galactic Center region, 
it is almost imperative to have a very accurate X-ray or radio position. This information
is today available, with sub-arcsec accuracy, through observations at radio wavelengths 
for both 
\unoe\ (Mirabel et al. 1992) 
and 
\grs\ (Mirabel \& Rodr\'{\i}guez 1993). On the other hand, it
is also required that the interstellar extinction does not completely prevent 
the observations.
In the \unoe\ case, there is little hope that an optical counterpart may be ever
found, given the strong absorption towards this source ($N_H\geq 8 \times 10^{22}$ cm$^{-2}$, 
Mirabel 1994). On the contrary, the absorption towards \grs\ is estimated to be
significantly lower, i.e, $N_H = (1.5\pm0.1)\times 10^{22}$ cm$^{-2}$ (Mereghetti et al. 1997a).
This corresponds to an extinction of $A_V\simeq8.4$ magnitudes in the optical and to 
only $A_K\simeq0.9$ magnitudes in the near infrared
(Rieke \& Lebofsky 1985; Predehl \& Schmitt 1995).
Thus, the search for an optical and near infrared counterpart to \grs\
should be regarded as a feasible project. Indeed, several observers have 
undertaken such a search (Mereghetti et al. 1992; Mirabel \& Duc 1992),  
but the initial lack of a sub-arcsec accurate radio position 
did not facilitate a successful result in this very crowded region. As a consequence, 
most of the subsequent discussions on the nature of \grs\   have been based on the 
bona fide assumption that only magnitude upper limits were available
(Chen et al. 1994; Mirabel 1994).

In this paper, we address again the issue of the \grs\ counterpart by re-analyzing
old images of the field as well as obtaining new deep ones. The astrometry has
been also revised and improved, with our attention being focused
on the Mirabel \& Rodr\'{\i}guez (1993) precise radio position, namely
R.A.(J2000)=18$^h$01$^m$12\rl 395  and DEC.(J2000)=$-25^{\circ}44^{\prime}$35\pri 90,
that is accurate to $\pm$0\pri 1.
As a result, we think now that there are a minimum of two possible
candidate counterparts for this microquasar source contrary to early
beliefs. The identification is mainly based at present on astrometric coincidence, 
and further observations would be necessary to confirm, or to rule
out, the proposed candidates. It is important to mention that one of them
is the same stellar-like object with $I\sim19$ originally pointed out 
by Mereghetti et al. (1994a).
The rest of the paper is devoted to discuss the possible stellar types
and luminosities of the normal companion of \grs\ that would be consistent
with the candidate counterparts.  Further details may be found 
in Chaty (1998).                           

\section{Observations and results}

Our observations were carried out using different telescopes at 
the European Southern Observatory\footnote{Based
on observations collected at the European Southern Observatory,
La Silla, Chile under proposal
numbers 59.D-0719, 60.D-0514 and 61.D-0431.} (ESO) in La Silla (Chile).
On several epochs between 1992 and 1997,
we obtained $J$, $H$ and $K$-band images with the IRAC2b camera at the ESO 2.2 m telescope.
For the epoch 1998 March 26, $UBVRI$ optical images of the region were similarly acquired 
using the ESO New Technology Telescope (NTT) with the EMMI CCD.
On 1998 May 18, NTT spectroscopy was also
obtained in the $H$ and $K$-bands with the SOFI infrared spectrograph and imaging camera.
All frames were reduced using standard procedures based on the IRAF image processing system.

\begin{figure}[htb]
\mbox{}
\vspace{10.5cm}
\includegraphics{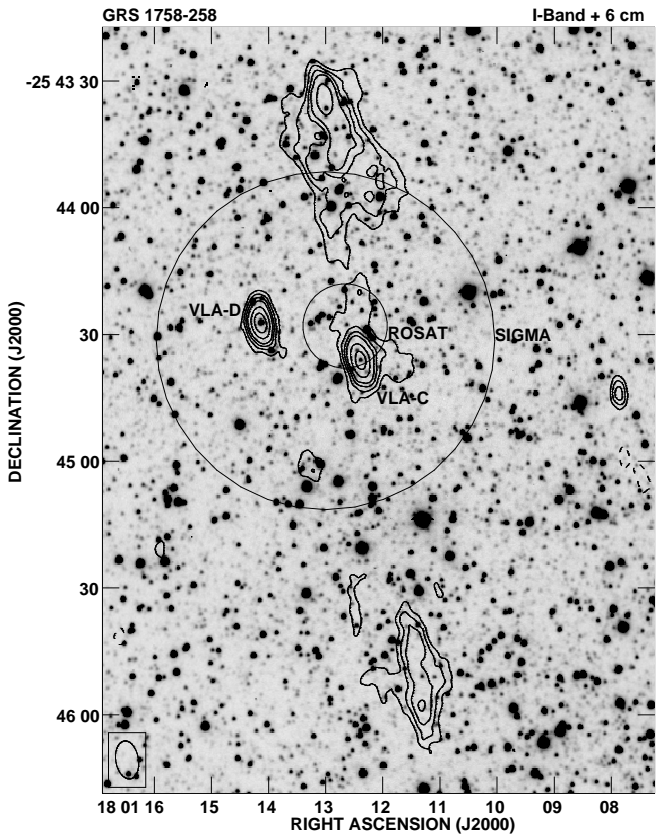}
\caption[]{I-band NTT image of the \grs\ field with VLA radio contours at 6 cm overlayed.
The SIGMA and ROSAT 90\% confidence error circles are also shown.
Their  radii are of $40^{\prime\prime}$  and $10^{\prime\prime}$, respectively (Goldwurm et al. 1994;
Mereghetti et al. 1994b).
Radio contours
are $-3$, 3, 4, 5, 6, 8 and 10 times 0.012 mJy beam$^{-1}$, the VLA map rms noise.
The corresponding synthesized beam is 9\pri 1$\times$6\pri 4, with position angle of 9\grp 5.}
\label{nttvla}
\end{figure}
 
\begin{figure}[htb]
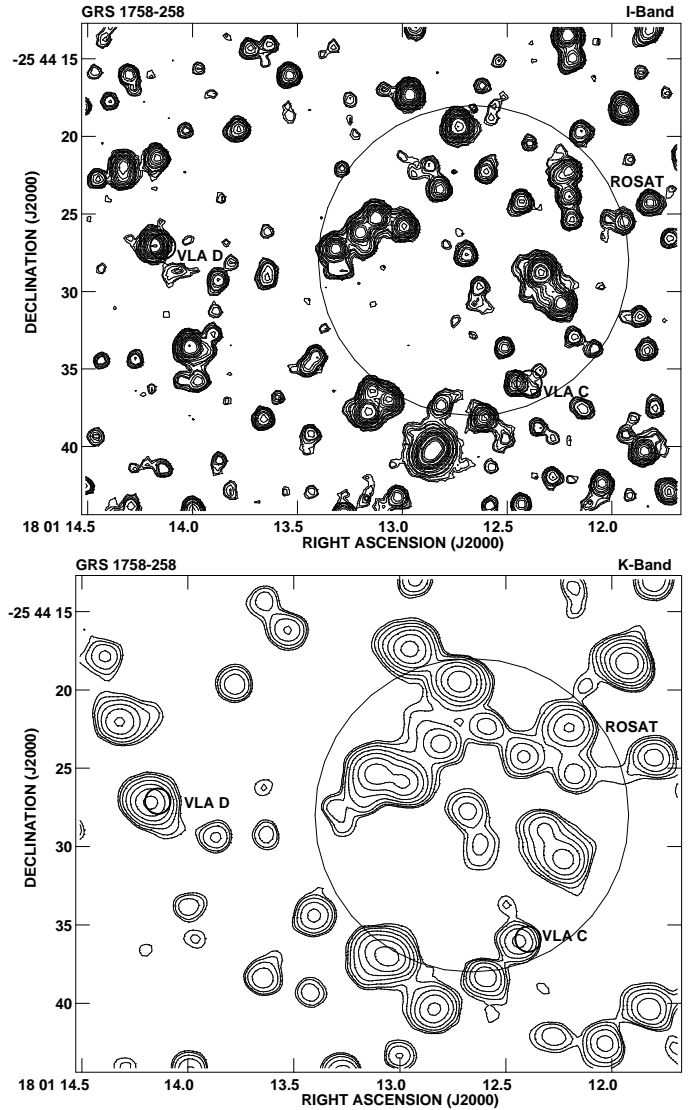

\mbox{}
\vspace{14.5cm}
\includegraphics{ag283.f2a}
\includegraphics{ag283.f2b}
\caption[]{Contour maps showing in detail the location of the \grs\ central source at
optical (NTT, I-band) and near infrared (2.2m, K-band) wavelengths. The 90\% confidence
error circle of the ROSAT position and that of the radio positions for VLA-C and VLA-D
are also marked.}
\label{ik}
\end{figure}

\subsection{Astrometry and photometry}

The astrometry on these images was obtained in two steps
involving the use of the Palomar Observatory Sky Survey
(POSS) digitized plates (Lasker et al. 1990).
First, we selected ten POSS stars in the GRS 1758$-$258 field
being included in the Hipparcos and Tycho catalogues (ESA, 1997). These objects were used as
primary reference stars in order to establish
a precise astrometric grid. This allowed us to measure the positions of six fainter stars
in the ESO frames being well distributed very close to the suspected position of \grs.
Afterwards, these fainter objects were adopted as secondary reference stars from
which the final astrometry could be determined.
We estimate that our total combined error is such that the $\pm$0\pri 1 radio positions
can be located with a 90\% confidence radius of about 0\pri 8 in the NTT
and 0\pri 9 in the 2.2 m frames, respectively.

\begin{table}
\caption[]{\label{mag} Magnitudes of the candidate counterparts}
\begin{tabular}{ccccc}
\hline
Filter   & Date of       &   1st VLA-C    & 2nd VLA-C     & VLA-D        \\
         & observation   &      Candidate &     Candidate & Candidate    \\
\hline
 $B$     & 1998 Mar 26   &  $>24.2$       &  $>24.2$      & $>24.2$      \\
 $V$     & 1998 Mar 26   &  $23.9\pm0.4$  &  $>24.9$      & $21.9\pm0.2$ \\
 $R$     & 1998 Mar 26   &  $21.0\pm0.2$  &  $22.6\pm0.3$ & $19.3\pm0.2$ \\
 $I$     & 1998 Mar 26   &  $19.0\pm0.2$  &  $21.1\pm0.3$ & $17.4\pm0.1$ \\
 $J$     &  1994-97      &  $16.2\pm0.1$  &    $-$        & $14.3\pm0.1$ \\
 $H$     &  1994-97      &  $14.8\pm0.1$  &    $-$        & $13.0\pm0.1$ \\
 $K$     &  1994-97      &  $14.0\pm0.1$  &    $-$        & $12.3\pm0.1$ \\
\hline
\end{tabular}
\end{table}

In Fig. \ref{nttvla} we show a wide field NTT image of the \grs\ region in the optical
I-band. The 90\% confidence error circles 
representing the SIGMA and ROSAT PSPC  
positions are plotted on it (Goldwurm et al. 1994; Mereghetti et al. 1994b). 
For illustration purposes, we have also overlayed  
the radio contours showing the extent  of the microquasar jets.
These contours correspond to a deep radio image obtained with the 
NRAO\footnote{The National Radio Astronomy Observatory is operated by
Associated Universities, Inc., under cooperative agreement with the
USA National Science Foundation.} Very Large Array (VLA), to be discussed in
more detail in a future paper. 
In addition to the jet extended emission, 
the \grs\ central component and another compact source to the east
are clearly detected. We will refer to them as sources VLA-C and VLA-D 
as in Rodr\'{\i}guez et al. (1992).

In Fig. \ref{ik} we present an enlarged view of the target position at optical ($I$-band)
and near infrared ($K$-band) wavelengths, in the form of contour maps. 
The location of radio sources VLA-C and VLA-D is indicated on these images by means
of their 90\% confidence circle. 
It is clear  that at least two optical objects (perhaps three) may be 
consistent with being the central source of the \grs\ radio counterpart or VLA-C. They
are   well detected individually only in  the NTT frames. 
In the infrared, the IRAC2b resolution is not good enough to resolve them.
A likely counterpart candidate to the radio source VLA-D
(R.A.(J2000)=18$^h$01$^m$14\rl 142 and DEC.(J2000)=$-25^{\circ}44^{\prime}$27\pri 08; $\pm$0\pri 1  
uncertainty)
is obvious in all ESO frames. The apparent Johnson magnitudes of all the optical and infrared sources relevant
to the following discussion are listed in Table \ref{mag}. Those of the second VLA-C
candidate may be partially contaminated by the proximity of the first brighter one. 
No significant photometric variations \simgt0.1 magnitudes have been detected on
time scales of weeks during follow-up infrared observations (Chaty 1998).  

Note that in some previous works    optical and infrared  magnitudes and upper limits  
were reported erroneously, and seem to imply long term variability when
compared with Table  \ref{mag}. The  VLA-C
candidate counterpart observed in May 1991 had $I\sim$19 (Mereghetti et al. 1997a, 1997b; 
the value   $I\sim$17  was previously reported owing to a  typing error in Mereghetti et al. 1994a).
The   upper limits of $K\sim$17 and or $I\sim$21 often quoted 
(e.g., Mirabel \&  Rodr\'{\i}guez 1993; Mirabel 1994; Chen et al. 1994)
resulted from preliminary radio positions and astrometry solutions  
and/or from misinterpretation of previously published results that unfortunately
propagated through the literature.
In conclusion, there is no evidence that any of the objects considered
as possible counterparts has varied.

\subsection{Infrared spectroscopy}

\begin{figure}[htb]
\mbox{}
\vspace{6.0cm}
\includegraphics{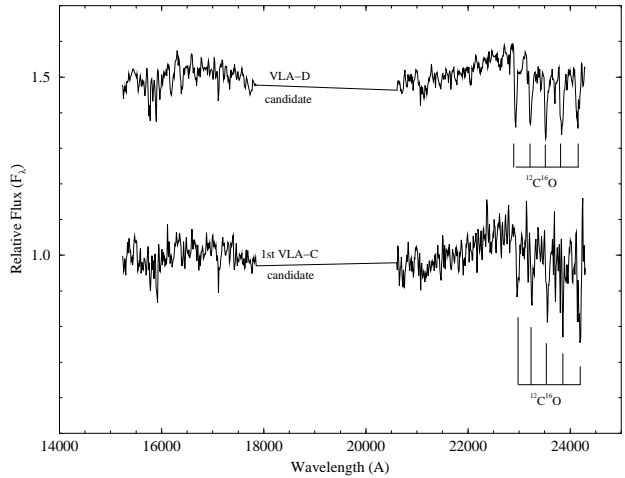}
\caption[]{Near infrared spectrum of the VLA-C and VLA-D  candidate counterparts obtained
with the SOFI instrument at the NTT on 1998 May 18. The continuum has been rectified
and normalized for better display.$^{12}$C$^{16}$O absorption band are the only
identified spectral features.
The huge gap between 1.8 and 2.1 $\mu$m
is due to the very strong atmospheric opacity.}
\label{sofi}
\end{figure}

The SOFI slit was aligned with a position angle
allowing to observe the VLA-C and VLA-D candidates simultaneously. 
The integration time amounted to one hour and consisted of short exposures
with the targets at different slit positions, thus allowing an accurate sky cancellation
during the processing. Atmospheric transmission was corrected using the Maiolino
et al. (1996) procedure. This involves division of the target spectrum by that of
a solar type standard, observed at the same air mass, plus a correction 
for the standard star absorption lines by means of a synthetic solar spectrum. 
The final result is shown in Fig. \ref{sofi}, where the main interesting
features are the clear $^{12}$C$^{16}$O bands in both the VLA-C first candidate
and in the VLA-D candidate as well. Feature identifications and equivalent widths are
listed in Table \ref{ew}.  
All the observed wavelengths appear close to the rest frame value, implying radial
velocities of at most \simlt300 km s$^{-1}$, i.e., as expected for a common stellar object
inside the Galaxy. The spectrum of the VLA-C second candidate is only very marginally
detected with SOFI and does not contaminate that of the first one.

\begin{table}
\caption[]{\label{ew} Equivalent widths of absorption features$^a$}
\begin{tabular}{cccc}
\hline
Feature               &   Wavelength & 1st VLA-C   &  VLA-D           \\
Identification        &   ($\mu$m)    & Candidate   & Candidate  \\
                      &              & (\AA)       & (\AA)         \\
\hline
$^{12}$C$^{16}$O(2,0) &   2.294      &    8     &  9 \\
$^{12}$C$^{16}$O(3,1) &   2.323      &    8     & 11 \\
$^{12}$C$^{16}$O(4,2) &   2.354      &   10     & 10 \\
$^{12}$C$^{16}$O(5,3) &   2.383      &    5     & 11 \\
$^{12}$C$^{16}$O(6,4) &   2.414      &   12     &  9 \\   
\hline
\end{tabular}

(a) Typical errors in the wavelength calibration are \\
$\pm0.001$ $\mu$m and $\pm1$ \AA\ in the equivalent width estimates.\\
\end{table}

\section{Discussion}

The fact that candidate counterparts are available now for
\grs\ opens the possibility to investigate the true nature of this
microquasar source, as well as to extrapolate the preliminary results to its
Galactic Center twin \unoe. In the following, we will always assume
that only VLA-C is associated with \grs\ and that VLA-D, being undetectable in X-rays
and misaligned with the jets, is most likely an unrelated source. 

The observed magnitudes in Table \ref{mag}, together with the recent
column density estimate $N_H = (1.5\pm0.1)\times 10^{22}$ cm$^{-2}$ by 
Mereghetti et al. (1997a), yield the following dereddened absolute
magnitudes assuming a 8.5 kpc distance. For the brighter VLA-C candidate, we find: 
$B>-1.5$, 
$V=+0.9\pm1.1$, 
$R=+0.1\pm0.6$, 
$I=+0.3\pm0.5$, 
$J=-0.8\pm0.3$, 
$H=-1.3\pm0.2$ and 
$K=-1.6\pm0.2$.
The corresponding dereddened colors are, e.g., 
$V-R=+0.8\pm1.7$, $V-I=+0.6\pm1.6$, $V-J=+1.7\pm1.4$, $V-H=+2.2\pm1.3$, $V-K=+2.5\pm1.3$,
$J-H=+0.5\pm0.5$, $J-K=+0.8\pm0.5$ and $H-K=+0.3\pm0.4$.
From these values, the most consistent spectral type would be an early K giant star
(Johnson 1966; Ruelas-Mayorga 1991). 
The agreement with an early K III companion is specially good for the absolute magnitudes.
The colors are less constraining due to errors, but nevertheless consistent with
this determination.
On the other hand, the presence of  $^{12}$C$^{16}$O absorption bands is
very typical of evolved late type stars (Kleinmann \& Hall 1986) thus giving
further support to our classification. 

If the association of this candidate with VLA-C is correct, \grs\
would be an X-ray binary system of low/intermediate mass. This suggests 
an interesting similarity of \grs\
with other persistent Galactic Bulge X-ray sources with weak radio emission, such as for instance 
GX 13+1 (Grindlay \& Seaquist 1986). 
The infrared spectrum of GX 13+1 also contains $^{12}$C$^{16}$O absorption
bands believed to be the signature of a late type giant companion (Bandyopadhyay
et al. 1997). However, the main problem with this interpretation is the absence
of observed emission lines. Emission lines should be normally
expected from the accretion disk of the system and they are not evident in the spectrum of
Fig. \ref{sofi} . In particular, we do not see any clearly detectable
Brackett-$\gamma$ emission, while other Galactic Bulge sources (e.g. Sco X-1, GX 13+1, GX 1+4)
do display it (Bandyopadhyay et al. 1997). Therefore, we cannot strictly rule out at present
that this counterpart candidate is a mere line-of-sight coincidence in
the crowded regions of the Galactic Bulge.
From the images in Fig. \ref{ik}, we find $\sim120$ $I$-band and $\sim50$ $K$-band
sources in a solid angle of 1000 arc sec$^2$. Then,
the a priori probability of finding an object in
our astrometric error circle is non negligible and amounts to $\sim20$\% and $\sim10$\%
in the optical and infrared, respectively.

Alternatively, for the weaker VLA-C counterpart candidate 
the approximate absolute magnitudes that we estimate
are $R=+1.7\pm0.9$ and $I=2.4\pm0.8$. With the same caution as above
for line-of-sight coincidences, these values would point towards
a main sequence F star as the most likely interpretation, thus 
implying a low-mass 
X-ray binary system. 
No reliable spectroscopic information is available for this source
given its weakness. 

The two VLA-C candidates reported here certainly exclude the possibility of a high mass 
companion for \grs.
It is also worth mentioning that, if we could see them 
through the same huge absorption as in \unoe, then their apparent
magnitudes would be consistent with the present \unoe\ limits. This fact is reassuring
that both microquasars are likely to have similar companions. 
Moreover, if the VLA-C first candidate is indeed \grs, its extrapolated apparent magnitude 
at $L'$-band (3.8$\mu$) assuming the \unoe\ absorption is found to be $L'$\simgt14.
This value is interestingly not far from the possible marginal 
detection of \unoe\ by Djorgovski et al. (1992) long ago ($L'=13\pm1$).    
In any case, we conclude this work by stressing the need for
further sensitive observations to confirm or to rule out the proposed identifications.


\begin{acknowledgements}
We thank the ESO staff for their valuable assistance
in obtaining and reducing some of the observations reported in this paper,
specially to Jean-Fran\c cois Gonz\'alez, Christopher Lidman and Fernando Comer\'on.
We also thank an anonymous referee for helping to improve this paper. 
\end{acknowledgements}

\end{document}